# Thermal properties and Magnetic susceptibility of Hellmann potential in Aharanov-Bohm(AB) flux and Magnetic fields at zero and finite temperatures

C. O. Edet[1*], P. O. Amadi[1], M. C. Onyeaju[1], U. S. Okorie[1&2], R. Sever[3], G. J. Rampho[4] and A. N. Ikot[1]

[1]Theoretical Physics Group, Department of Physics, University of Port Harcourt, Port Harcourt-Nigeria.

[2]Department of Physics, Akwa Ibom State University, Ikot Akpaden, Uyo.-Nigeria

[3]Department of Physics, Middle East Technical University, 06800, Ankara, Turkey.

[4]Department of Physics, University of South Africa.

## Abstract

In this research work, the Hellmann potential is studied in the presence of external magnetic and AB-flux fields. We solve the Schrodinger in the presence of these fields and the potential via the functional analysis approach (FAA). The energy equation and wave function of the system are obtained in closed form. The effect of the fields on the energy spectra of the system examined in details. It is found that the AB field performs better than the magnetic in its ability to remove degeneracy. Furthermore, the magnetization and magnetic susceptibility of the system was discussed at zero and finite temperatures. We evaluate the partition function and use it to evaluate other thermodynamic properties of the system such as magnetic susceptibility, $\chi_m(\vec{B}, \Phi_{AB}, \beta)$ Helmholtz free energy, $F(\vec{B}, \Phi_{AB}, \beta)$, entropy, $S(\vec{B}, \Phi_{AB}, \beta)$, internal energy, $U(\vec{B}, \Phi_{AB}, \beta)$, and specific heat, $C_v(\vec{B}, \Phi_{AB}, \beta)$. A comparative analysis of the magnetic susceptibility of the system at zero and finite temperature shows a similarity in the behavior of the system. A straight forward extension of our results to three dimension shows that the present result is consistent with what is obtains in literature.

## 1. INTRODUCTION

The Hellmann potential proposed by Han Hellman in 1935 is used as an approximation for the simplified description of complex systems. This potential arose in an attempt to replace the tedious effects of the motion of the core (i.e. non-valence) electrons of an atom and its nucleus with an effective potential [1-4]. Subsequently, several investigations was carried out with this potential; for instance, it has been applied to study bound state problems using different advanced mathematical techniques [5, 6]. The Hellmann potential have been put to use to study the approximate scattering state solutions in the relativistic regime [7-9]. The applications of this potential model include the following amongst many others; atomic physics and neutron scattering electron-core [10, 11], "electron-ion" [12], "inner-shell ionization problem", "alkali hydride molecules" and in condensed matter physics [13, 14]. The Hellmann potential is a superposition of the well-known Coulomb potential and the Yukawa (screened Coulomb) potential and it is expressed as [1, 5-6];

$$V(\rho) = -\frac{a}{\rho} + \frac{b}{\rho}e^{-\eta\rho} \tag{1}$$

where $a$ and $b$ are parameters that represent the strength of the Coulomb and Yukawa potentials respectively, $\eta$ represents the screening parameter and $\rho$ is the distance between the particles.

Based on many applicability of the Hellman potential, it will is essential to look into the bound state solutions of the two-dimensional (2D) non-relativistic (i.e. Schrödinger) wave equation with this potential under the collective influence of magnetic and Aharonov-Bohm (AB) fields. The solutions of the non-relativistic wave equation in two dimension with external fields has been a subject of great interest, as many researchers in the past have used this model to study many quantum mechanical phenomenon. For instance, Zakrzewski et al [15] studied the hydrogen atom model in two dimensions. The hydrogen atom under discussion was examined as an atomic spectroscopy and employed as an easy model for the ionization procedure which is extremely excited by circular-polarized microwaves. Eshghi et al[16] solved this equation in 2D with external magnetic and Aharonov–Bohm (AB) flux fields alongside a position-dependent mass (PDM) interacting with a superposed potential of "Morse" and "Coulomb" potentials respectively. The authors obtained the energy of the systems as well as their wave functions for two mass distribution functions that depends on position. Furthermore, the authors analyzed the thermal quantities for the system Again, Eshghi et al [17] solved this equation with a particle that is charged with mass function that is position-dependent in a Hulthen potential coupled with Coulomb-like potential field under the actions of the external magnetic and Aharonov–Bohm (AB) flux fields. The authors calculated bound state eigenvalues and eigenfunctions. Eshghi and Mehraban[18] also reported a general form of this equation in curved space by introducing Aharonov-Bohm (AB) flux and magnetic fields to the system. Subsequently, they solved the generalized model with the radial scalar power potential (RSPP) with the curvilinear coordinates system. The Pseudo-harmonic oscillator potential in the presence of magnetic field and Aharonov-Bohm (AB) flux field have been studied by Khordad [19], where he solved this non-relativistic equation exactly to obtain its bound states energy. Ikhdair and Falaye[20] solved the two-dimensional spinless Klein–Gordon (KG) equation with harmonic oscillator potential with and without magnetic and Aharonov–Bohm (AB) flux fields and the authors obtained exact energy eigenvalues and normalized wave functions. They went further to investigate the effects of these fields on the non-relativistic energy eigenvalues and wave functions obtained [20]. Ikhdair et al [21] solved the two-dimensional Schrödinger wave equation (SWE) with various power interaction potentials in the presence of magnetic and Aharonov-Bohm (AB) flux fields. The authors computed the energy levels of some diatomic molecules in the presence and absence of magnetic and AB flux fields using different quantum mechanical models. It was noted that the effect of the AB field is abundant as it makes a broader shift for $m \neq 0$ and its influence on $m = 0$ states was found to be superior to that of the magnetic field. Falaye et al[22] studied the effect of restraining the Hydrogen atom with the AB flux and electric and magnetic fields surrounded by quantum plasmas. The overall effects result in an intensely attractive system while the localizations of quantum levels change and the eigenvalues decreased accordingly. The authors found that the combined effect of the fields are much stronger than the isolated effect and consequently that there is a significant shift in the bound state energy of the system[22]. Aygun et al.[23] also solved the SWE in 2D solution for the Kratzer potential in the presence and absence of a constant magnetic field. The magnetic field effect on the energy spectra of the Kratzer potential was studied. Oyewumi et al [24] examined the effect of magnetic field on the bound state solution of the SWE with the pseudo-harmonic oscillator potential. It was discovered that the energy spectrum obtained mainly depends on dissociation energy and the magnetic quantum numbers $m$ which are influenced by the magnetic field. Ferkous and Bounames [25] solved the two dimensional Pauli equation with Hulthen potential for spin-1/2 particle in the presence of AB field. They obtained singular and regular solutions of the problem. It is shown that the AB field raises the degeneracy of the energy levels. Çetin [26] examined the effect of magnetic field on an electron that is free to move on a nanosphere. The exact energy levels and wave functions were also obtained. Landau energy levels was depicted for magnetic fields occurring on two-dimensional flat surfaces, when the radius is very large. In another interesting development, the Dirac-Weyl equation was used by Orozco et al [27] to find the exact energy equation of the graphene quantum dot interacting with AB-flux field and magnetic field. It was discovered that apart from using the graphene sheet and external magnetic field, the Aharonov-Bohm(AB)-flux field could as well be utilized to control the carriers state energies in graphene.

It is recognized that thermodynamics is branch of physics that offers analysis of macroscopic thermodynamic quantities at the molecular-level. It employs probability theory to the investigation of the thermodynamic activities systems comprised of a large number of particles. The elucidation of the macroscopic theory of thermodynamics in terms of the more abstract microscopic statistical mechanics was one of the most important triumphs of physics in the early twentieth century [28-30]. For a quantum system influenced by magnetic and AB field, some studies on thermodynamics properties have been carried out by a number of researchers among the selected few. Khordad and Sedehi[31] studied the thermodynamic properties of a Gallium Arsenide double ring shaped quantum dot in the presence of magnetic and electric fields. The radial part of the non-relativistic wave equation was solved with the ring shaped quantum. The obtained expressions for the energy equations and wave functions analytically. They calculated the entropy, heat capacity, average energy and magnetic susceptibility of the quantum dot in the presence of a magnetic field via the canonical ensemble approach. Sukirti et al. [32] examined the thermodynamic features of Rashba quantum dots with magnetic field. The thermodynamic properties of asymmetric parabolic quantum dot have been extensively treated by Ibragimov [33]. Again, Khordad and Sedeh[34] employed extensive and non-extensive entropies to study magnetic susceptibility of grapheme in non-commutative phase-space. From their results, it was found that the magnetic susceptibility has a positive value using the Shannon entropy. On the other hand, the authors obtained both positive and negative values for the magnetic susceptibility of graphene when Tsallis entropy was used. The magnetization and magnetic susceptibility of donor impurity in parabolic GaAs quantum dot have been studied by Alia et al[35] at finite temperature under the joint effect of external electric and magnetic fields. All the energy matrix elements were obtained analytically. Their computed results show that electric field can modify the magnetic properties of the QD GaAs medium by flipping the sign of its magnetic susceptibility from diamagnetic $(\chi_m < 0)$ to paramagnetic $(\chi_m > 0)$. Baghdasaryan et al[37] rewrote the magnetic field operator and the SWE in toroidal coordinates. This Hamiltonian operator in toroidal coordinates was used to evaluate the dependence of one-electron energy spectrum and wave function on the geometrical parameters of a toroidal quantum dot and magnetic field strength. The energy levels was used to evaluate the canonical partition function, which was used to obtain mean energy, heat capacity, entropy, magnetization, and susceptibility of non-interacting electron gas. Khordad et al [37] scrutinized the effect of a functional magnetic field on the entropy and internal energy of GaAs cylindrical quantum dot. For this reason, the Tsallis formalism is applied to obtain internal energy and entropy. It was noted that the entropy maximum increases with increasing dot radius and internal energy increases with increasing magnetic field.

In this investigation, our aim is in four fold, first we extend the works in refs [5, 6] and solve the SWE with the Hellmann potential model in the presence of external magnetic and AB flux fields. By using the functional analysis approach (FAA), we give detailed solutions of the 2D SWE with Hellmann potential models in the presence of magnetic and Aharonov-Bohm(AB) flux fields. The derived energy equation will be used to obtain the partition function which will in turn be used to obtain to other thermodynamic quantities like; entropy, mean free energy, specific heat capacity and magnetic susceptibility. We analyze the effect of the fields on these properties. More so, magnetization and magnetic susceptibility at zero temperature is considered as well.

The outline of our paper is as follows. In section 2, we give the solutions of the 2D Schrödinger equation with the Hellmann potential and vector potential $\vec{A}$ under the influence of external magnetic and AB flux fields. In section 3, the computations of numerical energy spectrum under external fields are considered and the comparison with previous results is given when fields become zero. In section 4, Magnetization and Magnetic susceptibility at Zero Temperature is considered. We study the behavior of thermodynamics properties in the presence of external fields in section 5. Finally, the paper ends with concluding remarks in section 6.

## 2. Schrödinger equation with Hellmann potential with AB flux and an external magnetic fields

The Hamiltonian operator of a particle that is charged and subjected to move in the Hellmann potential under the combined impact of AB flux and an external magnetic fields can be written in cylindrical coordinates. Thus, the SWE is written as in Ref. [16-18] taking into consideration the Hellmann potential.

$$\left(i\hbar\vec{\nabla}+\frac{e}{c}\vec{A}\right)^2\psi(\rho,\phi,z)=2\mu\left[E_{nm}+\frac{a}{\rho}-\frac{b}{\rho}e^{-\eta\rho}\right]\psi(\rho,\phi,z), \quad (2a)$$

where $E_{nm}$ denotes the energy level, $\mu$ is the effective mass of the system, the vector potential which is denoted by "$\vec{A}$" can be written as a superposition of two terms $\vec{A}=\vec{A_1}+\vec{A_2}$ having the azimuthal components [22] and external magnetic field with $\vec{\nabla}\times\vec{A_1}=\vec{B}, \vec{\nabla}\times\vec{A_2}=0$, where $\vec{B}$ is the magnetic field. $\vec{A_1}=\frac{\vec{B}e^{-\eta\rho}}{1-e^{-\eta\rho}}\phi$ and $\vec{A_2}=\frac{\phi_{AB}}{2\pi\rho}\phi$ represents the additional magnetic flux $\phi_{AB}$ created by a solenoid with $\vec{\nabla}\cdot\vec{A_2}=0$. The vector potential in full is written in a simple form as;

$$\vec{A}=\left(0,\frac{\vec{B}e^{-\eta\rho}}{1-e^{-\eta\rho}}+\frac{\phi_{AB}}{2\pi\rho},0\right).$$

$$\frac{1}{\rho^2}=\frac{\eta^2}{\left(1-e^{-\eta\rho}\right)^2} \quad (2b)$$

Let us take a wave function in the cylindrical coordinates as $\psi(\rho,\phi)=\frac{1}{\sqrt{2\pi\rho}}e^{im\phi}R_{nm}(\rho)$, where m denotes the magnetic quantum number. Inserting this wave function, the vector potential into Eq. (2) and using the approximation proposed by Greene and Aldrich [38] with some simple algebraic calculations, we arrive at the following radial second-order differential equation:

$$R''_{nm}(\rho) + \left[ \frac{2\mu E_{nm}}{\hbar^2} + \frac{2\mu}{\hbar^2}\left(\frac{\eta a}{(1-e^{-\eta\rho})} - \frac{\eta b}{(1-e^{-\eta\rho})}e^{-\eta\rho}\right) + \frac{2m\tau}{\hbar}\left(\frac{\eta \vec{B} e^{-\eta\rho}}{(1-e^{-\eta\rho})^2}\right) - \frac{\tau^2 \vec{B}^2 e^{-2\eta\rho}}{\hbar^2(1-e^{-\eta\rho})^2} - \frac{\tau^2 \eta \vec{B}\Phi_{AB} e^{-\eta\rho}}{\hbar^2(1-e^{-\eta\rho})^2 \pi} - \frac{\left[(m+\xi)^2 - \frac{1}{4}\right]\eta^2}{(1-e^{-\eta\rho})^2} \right] R_{nm}(\rho) = 0 \quad (3)$$

where $\tau = -\dfrac{e}{c}$, $\phi_0 = \dfrac{hc}{e}$ and $\xi = \dfrac{\Phi_{AB}}{\phi_0}$.

For Mathematical simplicity, let's introduce the following dimensionless notations;

$$-\varepsilon_{nm} = \frac{2\mu E_{nm}}{\hbar^2 \eta^2},\ \beta_1 = \frac{2\mu a}{\hbar^2 \eta},\ \beta_2 = \frac{2\mu b}{\hbar^2 \eta},\ \delta_1 = \frac{2m\tau \vec{B}}{\hbar\eta},\ \delta_2 = \frac{\tau^2 \vec{B}^2}{\hbar^2 \eta^2},\ \delta_3 = \frac{\tau^2 \vec{B}\Phi_{AB}}{\hbar^2 \eta \pi}\ \text{and}$$

$$\gamma = (m+\xi)^2 - \frac{1}{4} \quad (4)$$

Now using the functional analysis approach (FAA) [39] with the following substitution $s = e^{-\eta\rho}$ into Eq. (3), we can simply write Eq. (3) in the s-coordinate as follows;

$$\frac{d^2 R_{nm}(s)}{ds^2} + \frac{1}{s}\frac{dR_{nm}(s)}{ds} + \frac{1}{s^2(1-s)^2}\left[\begin{array}{l}-(\varepsilon_{nm} - \beta_2 + \delta_2)s^2 + (2\varepsilon_{nm} - \beta_1 - \beta_2 + \delta_1 - \delta_3) \\ -(\varepsilon_{nm} - \beta_1 + \gamma)\end{array}\right] R_{nm}(s) = 0 \quad (5)$$

If we consider the boundary conditions

$$s \Rightarrow \begin{cases} 0, & r \to \infty, \\ 1, & r \to 0, \end{cases}\ \text{when} \quad (6)$$

with $R(s) \to 0$, we take the following radial wave functions of the form

$$R(s) = s^\sigma (1-s)^\varphi f(s) \quad (7)$$

where

$$\sigma = \sqrt{\varepsilon_{nm} - \beta_1 + \gamma} \quad (8)$$

$$\varphi = \frac{1}{2} + \sqrt{\frac{1}{4} + \gamma - \delta_1 + \delta_3} \quad (9)$$

On substitution of Eq. (7) into Eq. (5) leads to the following hypergeometric equation:

$$s(1-s)f''(s) + \left[(2\sigma+1) - (2\sigma + 2\varphi + 1)s\right]f'(s) - \left[\left((\sigma+\varphi) - \sqrt{\varepsilon_{nm} - \beta_2 + \delta_2}\right)\left((\sigma+\varphi) + \sqrt{\varepsilon_{nm} - \beta_2 + \delta_2}\right)\right]f(s) = 0 \quad (10)$$

whose solutions are nothing but the hypergeometric functions

$$f(s) = {}_2F_1(a,b;c;s) \quad (11)$$

where
$$a = (\sigma+\varphi) - \sqrt{\varepsilon_{nm} - \beta_2 + \delta_2}$$
$$b = (\sigma+\varphi) + \sqrt{\varepsilon_{nm} - \beta_2 + \delta_2} \quad (12)$$
$$c = 2\sigma + 1$$

By considering the finiteness of the solutions, the quantum condition is given by
$$(\sigma+\varphi) - \sqrt{\varepsilon_{nm} - \beta_2 + \delta_2} = -n, \qquad n = 0,1,2... \quad (13)$$
from which we obtain

$$\varepsilon_{nm} = \beta_1 - \gamma + \left( \frac{\left(n + \frac{1}{2} + \sqrt{\frac{1}{4} + \gamma - \delta_1 + \delta_3}\right)^2 - \beta_1 + \gamma + \beta_2 - \delta_2}{2\left(n + \frac{1}{2} + \sqrt{\frac{1}{4} + \gamma - \delta_1 + \delta_3}\right)} \right)^2 \quad (14)$$

Hence, if one substitutes the value of the dimensionless parameters in Eq. (4) into Eq. (14), we obtain the solutions as follows:

$$E_{nm} = \frac{\hbar^2 \eta^2}{2\mu}\left((m+\xi)^2 - \frac{1}{4}\right) - \eta a - \frac{\hbar^2 \eta^2}{2\mu}\left(\frac{(n+\kappa)^2 - \frac{2\mu(a-b)}{\hbar^2 \eta} + (m+\xi)^2 - \frac{1}{4} - \frac{\tau^2 \vec{B}^2}{\hbar^2 \eta^2}}{2(n+\kappa)}\right)^2 \quad (15a)$$

where $\kappa = \frac{1}{2} + \sqrt{(m+\xi)^2 - \frac{2m\tau\vec{B}}{\hbar \eta} + \delta_3 = \frac{\tau^2 \vec{B} \Phi_{AB}}{\hbar^2 \eta \pi}}$, $m = \pm 1, \pm 2, \pm 3...$, and $m$ is the magnetic quantum number.

The three dimensional non-relativistic energy solutions are obtained by setting $m = \ell + \frac{1}{2}$ where $\ell$ is the rotational quantum number, in Eq. (15) to obtain;

$$E_{n\ell} = \frac{\hbar^2 \eta^2 \ell(\ell+1)}{2\mu} - \eta a - \frac{\hbar^2 \eta^2}{2\mu}\left(\frac{(n+\ell+1)^2 - \frac{2\mu(a-b)}{\hbar^2 \eta} + \ell(\ell+1)}{2(n+\ell+1)}\right)^2. \quad (15b)$$

Eq.(15b) is in excellent agreement with Eq.(33) of ref.[40] and refs.[41,42]

The corresponding unnormalized wave function is obtain as
$$R_{nm}(s) = N_{nm} s^{\sqrt{\varepsilon_{nm} - \beta_1 + \gamma}} (1-s)^{\frac{1}{2} + \sqrt{\frac{1}{4} + \gamma - \delta_1 + \delta_3}} {}_2F_1\left(-n, n + 2(\sigma+\varphi); 2\sigma+1; s\right) \quad (16)$$

## 3 Results and discussion

In table 1, we compute the energy eigenvalue using eq.(15a) for three cases when $\eta = 0.005$; when both fields are absent, degeneracy is present. By subjecting the system to only the magnetic field, the energy values are reduced and degeneracies are removed. The energy spectra become more negative and the system becomes strongly attractive as the quantum number $n$ increases for fixed $m$. When only the AB field is applied, the degeneracy are affected and the energy eigenvalues increases. The all-inclusive effect of the fields is stronger than the individual effects and consequently, there is a significant shift in the bound state energy of the system. In table 2, we compute the energy eigenvalue using eq.(15a) for three cases with an increased value of the screening parameter $(\eta = 0.01)$. When both fields are absent i.e. $B = 0, \Phi_{AB} = 0$, degeneracy is observed. Again, by exposing the Hellmann potential to only the magnetic field, the energy values are reduced and degeneracy are not affected. The energy levels become more negative and the system becomes more bounded as the quantum number $n$ increases for invariant $m$. When only AB flux is functional, the degeneracy is removed rapidly and energy eigenvalue increases. The overall effects indicate that the system is strongly attractive. Also, the joint effect of the external fields is stronger than the individual effects and consequently, there is a substantial changein the bound state energy of the system. Table 3 shows a comparison of the present result with results of other authors in three dimension using eq.(15b). It is noted here that our result is consistent with what obtains in literature.

In Fig. 1 we show the combined effect of the AB flux and magnetic fields on the energy values of the Hellmann potential. The confinement effect of the AB flux field on the quantum system is stronger than that of the magnetic field. This can be seen in Figs. 1(a) and 1(b) shows that the energy eigenvalue decreases as $\vec{B}$ increases. But the effect of the AB flux is seen as the energy increases with increasing value of $\Phi_{AB}$. In Fig. 2 we show the combined effect of the AB flux and magnetic fields on the energy values of the Hellmann potential. Again, the confinement effect of the AB flux field on the quantum system is stronger than that of the magnetic field. This can be seen in Figs. 2(a) and 2(b) shows that the energy eigenvalue increases as $\Phi_{AB}$ increases. But the effect of the magnetic field is seen as the energy decreases with increasing value of $\vec{B}$. The energy increases monotonically with increasing AB flux field in Fig.1(a), a similar behavior is observed in Fig.1(b) but the curve representing energy eigenvalue variation for $\vec{B} = 1T$ shows an invariant trend. Fig. (3) shows the variation of the magnetization against magnetic field a varied $\Phi_{AB}$. It is shown that the magnetization increases precipitously with increasing $\vec{B}$ but decreases for increasing values of $\Phi_{AB}$. From Fig.(4), it is observed that at zero temperature, the magnetic susceptibility of the quantum system is seen to be paramagnetic in the region of $\vec{B}$ values considered. $\chi_m(\vec{B}, \Phi_{AB})$ decreases with increasing values of $\Phi_{AB}$. More so, the relationship between $\chi_m(\vec{B}, \Phi_{AB}, \beta)$ and $\vec{B}$ is linear as $\chi_m(\vec{B}, \Phi_{AB}, \beta)$ increases linearly as $\vec{B}$ increases. From Fig. 5 it is shown that the partition function was almost constant but increased at $\vec{B} \approx 1.8T$ and decreased again at $\vec{B} \approx 2.4T$. Beyond this point, it remained constant throughout to $\vec{B} = 3T$. Furthermore, the partition function decreased as temperature value increased. It is seen that the partition function was pseudo constant as AB field increased but the three curves converged at $B \approx 3T$ and unanimous increase is observed in three curves. In the nearly constant region, the partition function was observed to be low as temperature upsurges but beyond this region, the partition function was found to increase with increasing values of temperature. Figs 7 and 8 shows that the partition function decreases as $\beta$ increases. In Fig.(9), the relationship between magnetization, $M(\vec{B}, \Phi_{AB}, \beta)$ and $\vec{B}$ shows a pseudo-sinusoid in the region $1 < \vec{B} < 3$. The magnetization rises and drops simultaneously for all three curves and rises again. This rise was continuous to $\vec{B} = 6T$. The magnetization increase with increasing values of temperature in the later region but the converse is observed in the former region. Magnetization, $M(\vec{B}, \Phi_{AB}, \beta)$ against AB flux field, $\Phi_{AB}$ with different $\beta$ is graphically displayed in fig.(10) shows that the Magnetization looks similar in the region $0 < \Phi_{AB} < 2$. Beyond this region, a sharp rise is observed. It is shown in figs 11 and 12 that the magnetization decreases as $\beta$ increases for varying $\vec{B}$ and $\Phi_{AB}$. In both cases, a sharp rise is noticed a certain value of $\beta$. This rise continues $\vec{B} = 0.01$ without drop but when $\vec{B} = 0.02$ and $\vec{B} = 0.03$, the trend remained unchanged. In Fig. 12, we notice a similar behavior when the representative curve for $\Phi_{AB} = 1$ shows a sharp rise at $\beta \approx 0.06$, this continues smoothly and drops again at $\beta \approx 0.13$. Fig. 13 reveals that as $\vec{B}$ increases, magnetic susceptibility, $\chi_m(\vec{B}, \Phi_{AB}, \beta)$ increases. If we monitor closely the variation of $\chi_m(\vec{B}, \Phi_{AB}, \beta)$ against $\vec{B}$ under different temperature conditions, it's observed that $\chi_m(\vec{B}, \Phi_{AB}, \beta)$ decreases with increasing $\beta$. The system also reveals some sort of saturation at large $\vec{B}$. The plot also shows that a paramagnetic $\left(\chi_m(\vec{B}, \Phi_{AB}, \beta) > 0\right)$ behavior is dominant in the system over a range of $\vec{B}$. This is similar to the behavior of the quantum system at zero temperature. A closer look at the curves for $\beta = 0.04$ and $\beta = 0.08$, it is seen that in the region where $0.2T < \vec{B} < 1.1T$, the susceptibility was quasi constant but increased swiftly from $\vec{B} = 1.1T$. Fig.14 shows that as $\Phi_{AB}$ increases, magnetic susceptibility, $\chi_m(\vec{B}, \Phi_{AB}, \beta)$ decreases monotonically. The susceptibility increases for increasing values of $\beta$. The variation of the magnetic susceptibility with the AB flux field shows a diamagnetic behavior for $\beta = 0.04$ and $\beta = 0.08$. The magnetic susceptibility shows a slight paramagnetic behavior for $\beta = 0.01$. Fig. 15shows that as $\beta$ increases, magnetic susceptibility, $\chi_m(\vec{B}, \Phi_{AB}, \beta)$ increases monotonically. The susceptibility increases for increasing values of $\vec{B}$. The variation of the magnetic susceptibility with $\beta$ shows a diamagnetic behavior for varying $\vec{B}$. Fig. 16 plot shows that as $\beta$ increases, magnetic susceptibility, $\chi_m(\vec{B}, \Phi_{AB}, \beta)$ increases monotonically. The susceptibility decreases for increasing values of $\Phi_{AB}$. The variation of the magnetic susceptibility with $\beta$ shows a diamagnetic behavior for varying $\Phi_{AB}$. Fig.(17) shows the variation of the internal energy, $U(\vec{B}, \Phi_{AB}, \beta)$ with increasing magnetic field. The internal energy reduces for increasing values of $\beta$. It is observed that the $U(\vec{B}, \Phi_{AB}, \beta)$ decreases with increasing $\vec{B}$. We also notice a uniform drop of $U(\vec{B}, \Phi_{AB}, \beta)$ in all three cases at $\vec{B} = 2T$ and sharp rise. Fig.(18) shows the internal energy variation with AB flux field at varied $\beta$. The internal energy decreases with increasing $\Phi_{AB}$, and also increases with increased value of $\beta$. We also notice that the three curves converge at $\Phi_{AB} = 2.5$ for all $\beta$ values. Fig. 19 shows the internal energy, $U(\vec{B}, \Phi_{AB}, \beta)$ against $\beta$ with varied $\vec{B}$. The internal energy increases with increasing $\beta$ but drops and remains unchanged up to $\beta \approx 0.20$. This behavior is evident in three cases of $\vec{B}$. The internal energy, $U(\vec{B}, \Phi_{AB}, \beta)$ is plotted against $\beta$ with varied $\Phi_{AB}$ in Fig. (20). The internal energy of the system decreases with increasing $\beta$. Fig. 21 shows the variation of specific of heat, $C_v(\vec{B}, \Phi_{AB}, \beta)$ with $\vec{B}$ at varied temperature. The specific heat capacity increases with increasing $\vec{B}$, up to $\vec{B} \approx 2T$, beyond this point the specific heat drops at $\vec{B} = 4T$, $\vec{B} = 4.5T$ and $\vec{B} = 6T$ for $\beta = 0.08, \beta = 0.04$ and $\beta = 0.01$ respectively. After this point, the specific rises again and maintains a constant trend. Fig. (22) shows the variation of specific heat capacity, $C_v(\vec{B}, \Phi_{AB}, \beta)$ against $\Phi_{AB}$ with varied $\beta$. The specific heat capacity lowers as AB flux field increases. We also note that there is a sharp rise at $\Phi_{AB} \simeq 2$, afterwards the decrease was continuous. Fig. (23) shows the specific heat capacity, $C_v(\vec{B}, \Phi_{AB}, \beta)$ with varying $\beta$ and $\vec{B}$. The specific heat capacity decreases with increasing $\beta$, for $\vec{B} = 0.02T$ and $\vec{B} = 0.03$, although the variation shows a rise and low pattern. On the other hand, the curve for $\vec{B} = 0.01T$ shows a rising trend up to $\beta \simeq 0.14$ where it drops continuously to $\beta \simeq 0.17$. Fig.(24) shows the variation of the specific heat capacity with $\beta$, with varied values of $\Phi_{AB}$. $C_v(\vec{B}, \Phi_{AB}, \beta)$ decreases with increasing $\beta$, although it shows a rise and low nature in the trend. For $\Phi_{AB} = 1$ and $\Phi_{AB} = 2$, the specific heat drops at $\beta \simeq 0.05$ and $\beta = 0.06$ respectively and increases very slightly and drops again immediately. Thereafter, a constant trend is maintained. When $\Phi = 3$, the specific heat capacity peaks at $\beta = 0.05$ and then drops immediately to its minimum at $C_v(\vec{B}, \Phi_{AB}, \beta) = -1.2 \dfrac{J}{K}$ at $\beta = 0.11$ and then rise again and drops at $\beta \simeq 0.17$. Fig. 25 shows the plot of free energy, $F(\vec{B}, \Phi_{AB}, \beta)$ against $\vec{B}$ varying $\beta$. When $\beta = 0.01$, the free energy is higher, we observe that the free energy increases with increasing magnetic field and reaches it maximum at $\vec{B} \approx 0.04T$, from then on it falls sporadically. A similar trend is observed when $\beta = 0.04$ and $\beta = 0.08$, it rises at $\vec{B} \approx 0.03T$ and drops also. Fig. 25 shows the plot of Free energy, $F(\vec{B}, \Phi_{AB}, \beta)$ against $\Phi_{AB}$ varying $\beta$. When $\beta = 0.01$, again, the free energy is higher, we observe that the free energy increases with increasing magnetic field and reaches it maximum at $\Phi_{AB} \approx 2.4$, from then on it falls sporadically. A similar trend is observed when $\beta = 0.04$ and $\beta = 0.08$, it rises at $\Phi_{AB} \approx 2.5$ and drops also. In Fig. (27), the free energy, $F(\vec{B}, \Phi_{AB}, \beta)$ is plotted against $\beta$ with varying $\vec{B}$. It is observed that the free energy increases at a monotonic pattern as $\beta$ increases for all values of $\vec{B}$. The free energy, $F(\vec{B}, \Phi_{AB}, \beta)$ is plotted against $\beta$ with varying $\Phi_{AB}$ in Fig.(28). Once again, It is observed that the free energy increases at a monotonic pattern as $\beta$ increases for all values of $\Phi_{AB}$. We also notice that the higher the AB flux field, the lower the free energy. The entropy of the quantum mechanical system against the external magnetic field with varying $\beta$ is depicted in Fig.(29), the entropy of the system reduces as $\vec{B}$ increases up to $\vec{B} \approx 0.004T$ and immediately rises again for all three values of $\beta$.

Again, the entropy of the system against the AB-flux field with varying $\beta$ is plotted in Fig.(30), the entropy of the system diminishes as $\Phi_{AB}$ up to $\Phi_{AB} \approx 2.6$ and immediately rises again for all three values of $\beta$. It peaks up for $\beta = 0.01$ but a sharp decrease is observed for $\beta = 0.04$ and $\beta = 0.08$. Fig. (31) shows the entropy, $S(\vec{B}, \Phi_{AB}, \beta)$ with varying $\beta$ and $\vec{B}$. $S(\vec{B}, \Phi_{AB}, \beta)$ decreases with increasing $\beta$, for $\vec{B} = 0.02T$ and $\vec{B} = 0.03T$. On the other hand, the curve for $\vec{B} = 0.01T$ shows a rising trend up to $\beta \simeq 0.14$ where it drops continuously to $\beta \simeq 0.20$. Figs 32 shows the variation of entropy, $S(\vec{B}, \Phi_{AB}, \beta)$ with $\beta$ with varying $\Phi_{AB}$. The entropy decreases as $\beta$ increases but and decreases also for $\Phi_{AB}$.

**Table1:** Energy values for the Hellmann potential model under the influence of AB flux and external magnetic fields with various values of magnetic quantum numbers. The following fitting parameters have been employed: $\hbar = b = \mu = \phi = e = c = 1, a = 2$ and $\eta = 0.005$. All values are in natural units.

| m | n | $B=0, \Phi_{AB}=0$ | $B=5, \Phi_{AB}=0$ | $B=0, \Phi_{AB}=5$ | $B=5, \Phi_{AB}=5$ |
|---|---|---|---|---|---|
| 0 | 0 | -2.010003125 | -12510002.01 | -0.021986596 | -1881.110061 |
|   | 1 | -0.230008681 | -1389994.674 | -0.017892474 | -1791.688849 |
|   | 2 | -0.087621125 | -500394.0875 | -0.015344014 | -1708.475408 |
|   | 3 | -0.048407207 | -255300.0486 | -0.013661603 | -1630.908157 |
| 1 | 0 | -0.228892014 | -1522.367263 | -0.01722206 | -841.7223671 |
|   | 1 | -0.087215125 | -1456.955969 | -0.014823347 | -814.4719178 |
|   | 2 | -0.048197003 | -1395.655192 | -0.013241014 | -788.5143204 |
|   | 3 | -0.032157446 | -1338.127599 | -0.012152709 | -763.7690685 |
| -1 | 0 | -0.228892014 | 1570.192051 | -0.030250039 | 7959.937689 |
|   | 1 | -0.087215125 | 1565.505879 | -0.022748993 | 7839.387741 |
|   | 2 | -0.048197003 | 1556.180231 | -0.018454308 | 7604.307172 |
|   | 3 | -0.032157446 | 1542.307525 | -0.015780014 | 7266.113808 |

**Table2:** Energy values for the Hellmann potential model under the influence of AB flux and external magnetic fields with various values of magnetic quantum numbers. The following fitting parameters have been employed: $\hbar = b = \mu = \phi = e = c = 1, a = 2$ and $\eta = 0.01$. All values are in natural units.

| m | n | $B=0, \Phi_{AB}=0$ | $B=5, \Phi_{AB}=0$ | $B=0, \Phi_{AB}=5$ | $B=5, \Phi_{AB}=5$ |
|---|---|---|---|---|---|
| 0 | 0 | -2.0200125 | -3130002.02 | -0.027450517 | -914.5349707 |
|   | 1 | -0.2378125 | -347772.46 | -0.023995932 | -854.4683858 |
|   | 2 | -0.0952845 | -125194.0953 | -0.021909389 | -800.0935723 |
|   | 3 | -0.056077806 | -63871.48466 | -0.020597967 | -750.7134985 |
| 1 | 0 | -0.235568056 | -751.357965 | -0.035814969 | 4146.443075 |
|   | 1 | -0.0944605 | -706.3267065 | -0.0290125 | 4016.824466 |
|   | 2 | -0.055645153 | -665.194507 | -0.02517818 | 3770.791577 |
|   | 3 | -0.039740895 | -627.523898 | -0.022853389 | 3432.021333 |
| -1 | 0 | -0.235568056 | -751.357965 | -0.02261605 | -411.0410648 |
|   | 1 | -0.0944605 | -706.3267065 | -0.0208045 | -392.4168263 |
|   | 2 | -0.055645153 | -665.194507 | -0.01967686 | -375.0124326 |
|   | 3 | -0.039740895 | -627.523898 | -0.018970949 | -358.7236448 |

**Table 3**: Comparison of energy spectrum obtained from FAA with SUSY, Nikiforov-Uvarov (NU) and Amplitude Phase method with $\hbar = b = 2\mu = 1$ and $a = 4\mu$ [..].

| State | | present | SUSY[5] | pNU[6] | APM[6] |
|---|---|---|---|---|---|
| 1s | 0.001 | -0.251500250 | -0.251 500 | -0.251 500 | -0.250 969 |
|    | 0.005 | -0.257506250 | -0.257 506 | -0.257 506 | -0.254 933 |
|    | 0.01  | -0.265025000 | -0.265 025 | -0.265 025 | -0.259 823 |
| 2s | 0.001 | -0.064001000 | -0.064 001 | -0.064 001 | -0.063 243 |
|    | 0.005 | -0.070025000 | -0.070 025 | -0.070 025 | -0.067 106 |
|    | 0.01  | -0.077600000 | -0.077 600 | -0.077 600 | -0.071 689 |
| 2p | 0.001 | -0.064250250 | -0.063 750 | -0.064 000 | -0.063 495 |
|    | 0.005 | -0.071256250 | -0.068 756 | -0.070 000 | -0.067 377 |
|    | 0.01  | -0.080025000 | -0.075 025 | -0.077 500 | -0.072 020 |
| 3s | 0.001 | -0.029280028 | -0.029 280 | -0.029 280 | -0.028 283 |
|    | 0.005 | -0.035334028 | -0.035 334 | -0.035 334 | -0.031 993 |
|    | 0.01  | -0.043002778 | -0.043 003 | -0.043 003 | -0.036 142 |
| 3p | 0.001 | -0.029390250 | -0.029 169 | -0.029 279 | -0.028 765 |
|    | 0.005 | -0.035867361 | -0.034 756 | -0.035 309 | -0.032 480 |
|    | 0.01  | -0.044025000 | -0.041 803 | -0.042 903 | -0.036 142 |
| 3d | 0.001 | -0.029611361 | -0.028 945 | -0.029 388 | -0.028 767 |
|    | 0.005 | -0.036950694 | -0.033 617 | -0.035 817 | -0.032 526 |
|    | 0.01  | -0.046136111 | -0.039 469 | -0.043 825 | -0.036 613 |
| 4s | 0.001 | -0.017129000 | -0.017 129 | -0.029 280 | -0.016 601 |
|    | 0.005 | -0.023225000 | -0.023 225 | -0.035 334 | -0.020 077 |
|    | 0.01  | -0.031025000 | -0.031 025 | -0.043 003 | -0.023 551 |
| 4p | 0.001 | -0.017190563 | -0.017 066 | -0.017 128 | -0.016 602 |
|    | 0.005 | -0.023514063 | -0.022 889 | -0.023 200 | -0.020 098 |
|    | 0.01  | -0.031556250 | -0.030 306 | -0.030 925 | -0.023 641 |
| 4d | 0.001 | -0.017314063 | -0.016 939 | -0.017 189 | -0.016 604 |
|    | 0.005 | -0.024101563 | -0.022 227 | -0.023 464 | -0.020 098 |
|    | 0.01  | -0.032656250 | -0.028 906 | -0.031 356 | -0.023 641 |
| 4f | 0.001 | -0.017500250 | -0.016 750 | -0.017 311 | -0.016 607 |
|    | 0.005 | -0.025006250 | -0.021 257 | -0.024 024 | -0.020 142 |
|    | 0.01  | -0.034400000 | -0.026 900 | -0.032 356 | -0.024 056 |

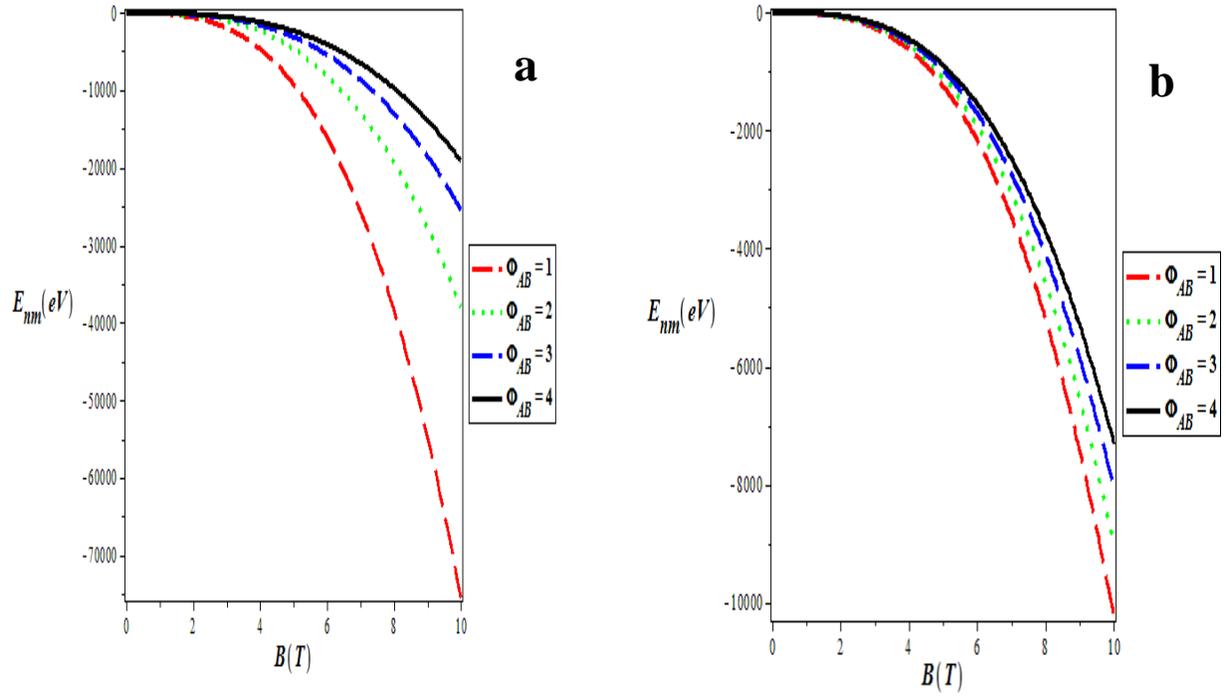

Figure 1: Variation of energy values for the Hellmann potential and under the influence of the magnetic field and the AB flux field in natural units using the fitting parameters $\hbar = b = \mu = \phi = e = c = 1, a = 2$ and $\eta = 0.005$ (a) as a function of external magnetic field with various $\Phi_{AB}$ and $m = n = 0$. (b) Same as (a) but with $m = n = 1$.

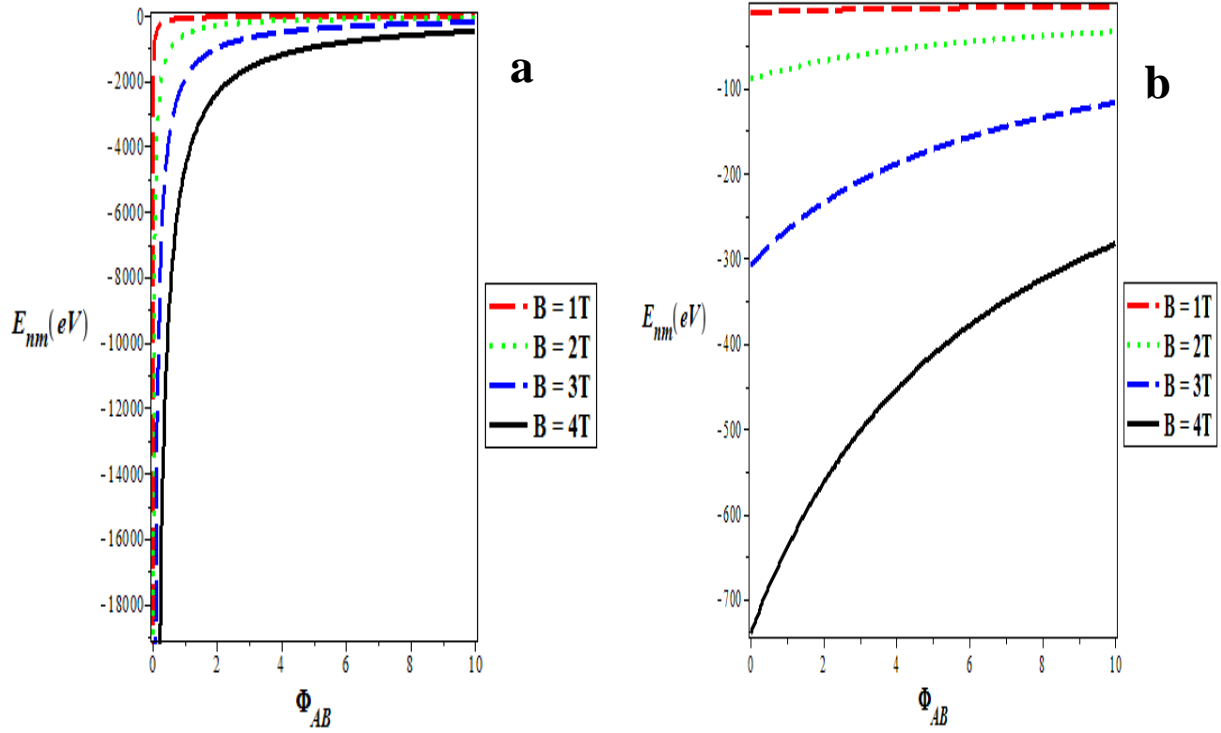

Figure 2: Variation of energy values for the Hellmann potential and under the influence of the magnetic field and the AB flux field in natural units using the fitting parameters $\hbar = b = \mu = \phi = e = c = 1, a = 2$ and $\eta = 0.005$ (a) as a function of AB flux field with various $\vec{B}$ and $m = n = 0$. (b) Same as (a) but with $m = n = 1$.

## 4 Magnetization and Magnetic susceptibility at Zero Temperature.

In the present study, we are interested in analyzing the magnetization and magnetic susceptibility at zero temperature.

### 4.1 Magnetization

The magnetization of a system in a state $(n, m)$ are defined by[43];

$$M_{nm}\left(\vec{B}, \Phi_{AB}\right) = -\frac{\partial E_{nm}}{\partial \vec{B}} \qquad (17)$$

### 4.2 Magnetic Susceptibility at zero temperature

The magnetic susceptibility at zero temperature is given as[43];

$$\chi_m = \frac{\partial M}{\partial \vec{B}} \qquad (18)$$

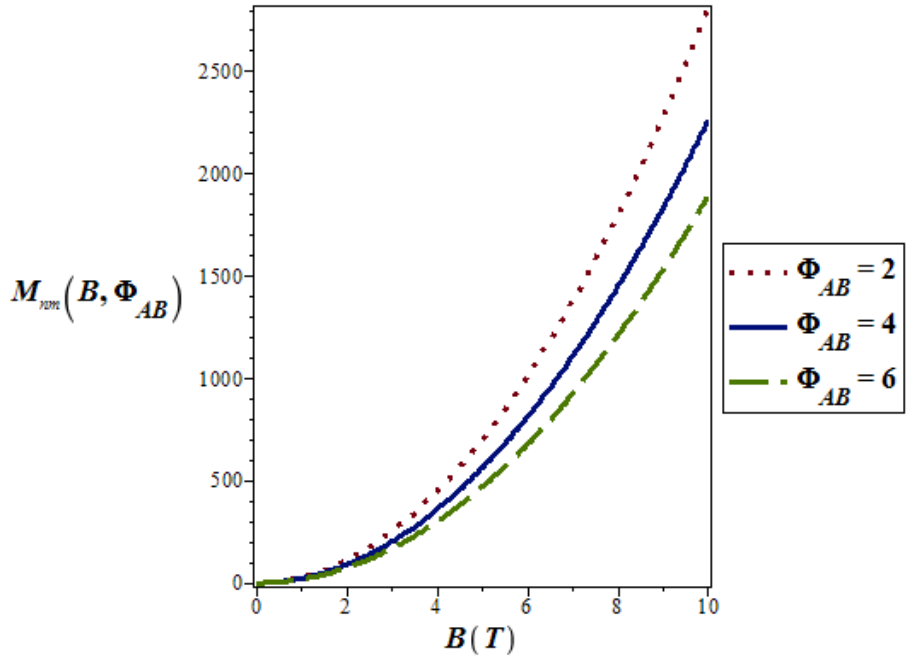

**Figure 3**: Plot of Magnetization against magnetic field for different values of AB flux field at zero temperature.

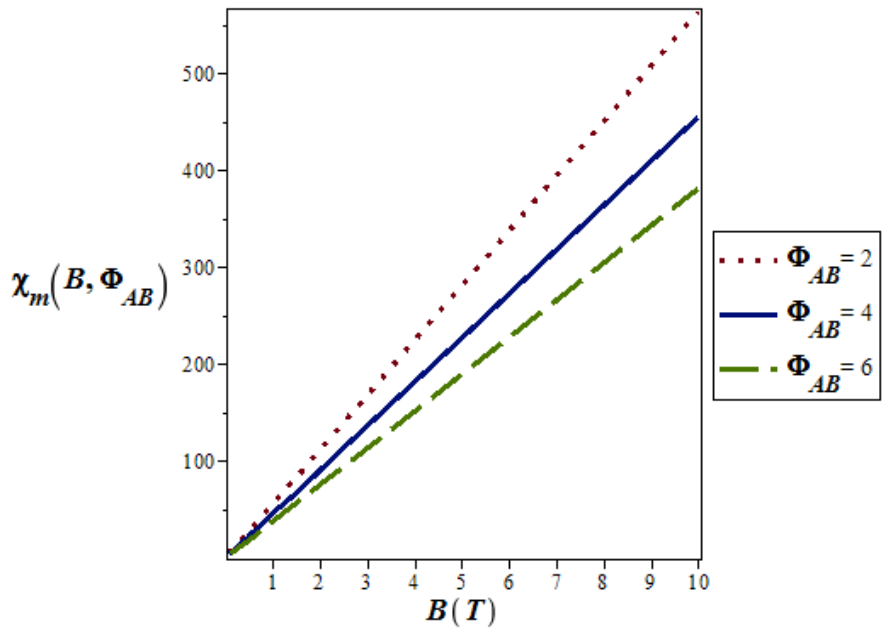

**Figure 4**: Plot of Magnetic Susceptibility against magnetic field for different values of AB flux field at zero temperature.

## 5  Thermal Properties of Hellmann Potential with Magnetic and AB fields

The vibrational partition function can be calculated with the aid of direct summation over all possible vibrational energy levels at a given temperature $T$ to be [44-46]

$$Z(\beta) = \sum_{n=0}^{\varpi} e^{-\beta E_{nm}}, \beta = \frac{1}{k_B T} \tag{19}$$

Here, $k_B$ is the Boltzmann constant and $E_{nm}$ is energy of the nth bound state.

We can rewrite eq. (15) to be of the form

$$E_{nm} = \Lambda_1 - \frac{\hbar^2 \eta^2}{2\mu} \left( \frac{(n+\kappa)^2 + \Lambda_2}{2(n+\kappa)} \right)^2 \tag{20}$$

$$\Lambda_1 = \frac{\hbar^2 \eta^2}{2\mu} \left( (m+\xi)^2 - \frac{1}{4} \right) - \eta a; \quad \Lambda_2 = -\frac{2\mu(a-b)}{\hbar^2 \eta} + (m+\xi)^2 - \frac{1}{4} - \frac{\tau^2 \vec{B}^2}{\hbar^2 \eta^2} \tag{21}$$

We substitute eq. (20) into eq. (19) to have

$$Z(\beta) = \sum_{n=0}^{\varpi} e^{-\beta \left[ \Lambda_1 - \frac{\hbar^2 \eta^2}{2\mu} \left( \frac{(n+\kappa)^2 + \Lambda_2}{2(n+\kappa)} \right)^2 \right]} \tag{22}$$

where, $\varpi = -\kappa + \sqrt{\Lambda_1} \pm \sqrt{\Lambda_1 - \Lambda_2}$ (23)

is the maximum quantum number.

In the classical limit, the sum in Eq.(22) can be replaced by an integral, such that

$$Z(\beta) = \int_0^{\varpi} e^{\beta \left( P(n+\kappa)^2 + \frac{Q}{(n+\kappa)^2} + R \right)} dn \tag{24}$$

where

$$P = \frac{\hbar^2 \eta^2}{8\mu}; Q = \frac{\hbar^2 \eta^2 \Lambda_2^2}{8\mu}; R = \frac{\hbar^2 \eta^2 \Lambda_2}{2\mu} + \Lambda_1. \tag{25}$$

$$Z(\beta) = \int_{\kappa}^{\varpi+\kappa} e^{-\beta \left( \frac{a}{\psi^2} + b\psi^2 + c \right)} d\psi, \psi = n + \kappa \tag{26}$$

The integral is evaluated in the region $\kappa \leq \psi \leq \varpi + \kappa$

We therefore use the Mathematica software to evaluate the integral in eq. (26), thus obtaining the partition function for the Hellmann potential model as;

$$Z(\vec{B},\beta,\Phi_{AB}) = \frac{e^{R\beta-2\sqrt{-P\beta}\sqrt{-Q\beta}}\pi\left(-1+Erf[v-\theta]+e^{4\sqrt{-P\beta}\sqrt{-Q\beta}}\left(-Erf[v+\theta]+Erf[\zeta+\varsigma]\right)+Erfc[\zeta-\varsigma]\right)}{4\sqrt{-Q\beta}} \quad (27)$$

where we have also introduced the following parameters for mathematical simplicity,

$$v = \frac{\sqrt{-P\beta}}{\varpi}, \zeta = \frac{\sqrt{-P\beta}}{\kappa+\varpi}, \theta = \sqrt{-Q\beta}\varpi \text{ and } \varsigma = \sqrt{-Q\beta}(\kappa+\varpi) \quad (28)$$

The error function can be defined as [47]

$$erf(z) = \frac{2}{\sqrt{\pi}}\int_0^z e^{t^2} dt \quad (29)$$

Thermodynamic functions such as; Magnetic Susceptibility, $\chi_m(\vec{B},\Phi_{AB},\beta)$ Helmholtz free energy, $F(\vec{B},\Phi_{AB},\beta)$, entropy, $S(\vec{B},\Phi_{AB},\beta)$, internal energy, $U(\vec{B},\Phi_{AB},\beta)$, and specific heat, $C_v(\vec{B},\Phi_{AB},\beta)$, functions can be obtained from the partition function(30) as follows;

**Magnetization at Finite Temperature**

The magnetization is given as[48];

$$M(\vec{B},\Phi_{AB},\beta) = \frac{1}{\beta}\left(\frac{1}{Z(\vec{B},\Phi_{AB},\beta)}\right)\left(\frac{\partial}{\partial \vec{B}}Z(\vec{B},\Phi_{AB},\beta)\right) \quad (30)$$

**Magnetic Susceptibility**;

The magnetic susceptibility of the system is calculated with [48]

$$\chi_m(\vec{B},\Phi_{AB},\beta) = \frac{\partial M(\vec{B},\Phi_{AB},\beta)}{\partial \vec{B}} \quad (31)$$

**Internal Energy**

The internal energy of the system is obtained as [49];

$$U(\vec{B},\Phi_{AB},\beta) = -\frac{\partial\left(\ln Z(\vec{B},\Phi_{AB},\beta)\right)}{\partial \beta} \quad (32)$$

**Specific Heat Capacity**

The specific heat capacity of is evaluated using the equation [49];

$$C_v\left(\vec{B}, \Phi_{AB}, \beta\right) = k_\beta \frac{\partial U\left(\vec{B}, \Phi_{AB}, \beta\right)}{\partial \beta} \tag{33}$$

**Free Energy**

The free energy of the system is given as [49]

$$F\left(\vec{B}, \Phi_{AB}, \beta\right) = -\frac{1}{\beta} \ln Z\left(\vec{B}, \Phi_{AB}, \beta\right) \tag{34}$$

**Entropy**

The entropy of the system is evaluated with the expression below[49];

$$S\left(\vec{B}, \Phi_{AB}, \beta\right) = -k_\beta \frac{\partial F\left(\vec{B}, \Phi_{AB}, \beta\right)}{\partial \beta} \tag{35}$$

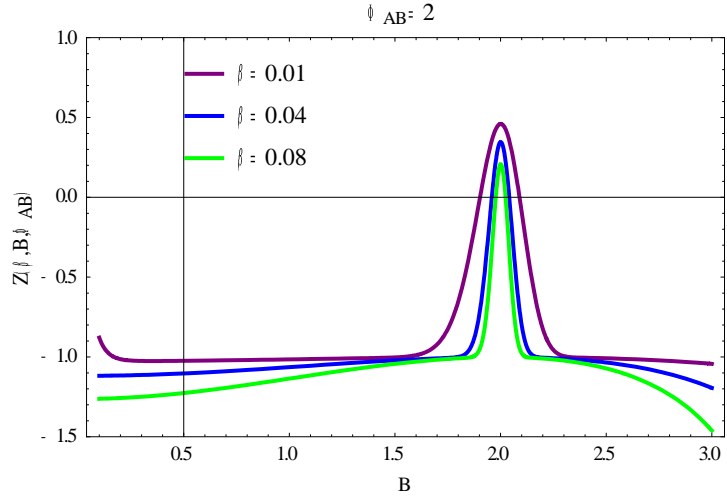

**Figure 5**: Plot of Partition function against magnetic field for different values of temperature.

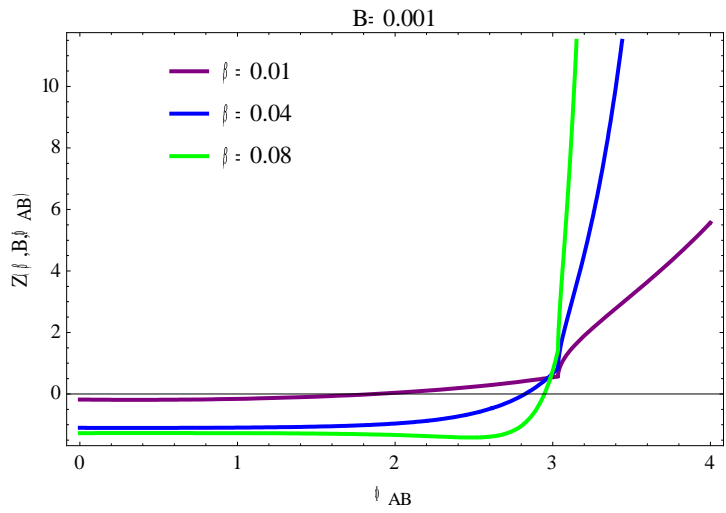

**Figure 6**: Plot of Partition function against AB flux field for different values of temperature.

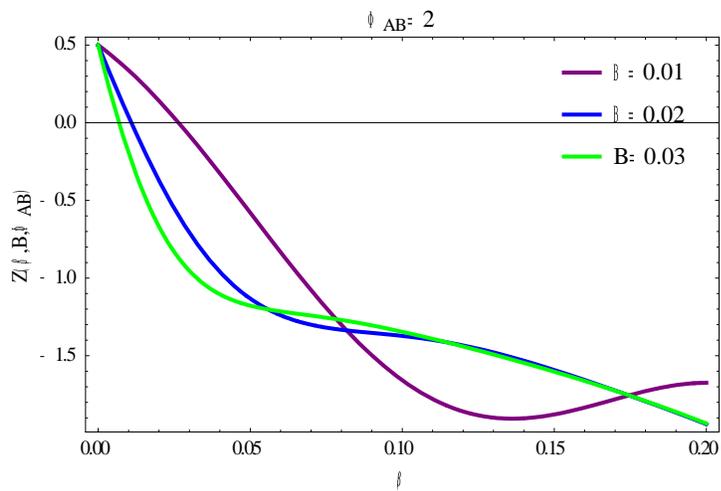

**Figure 7**: Plot of Partition function against $\beta$ for different values magnetic field, $\vec{B}$.

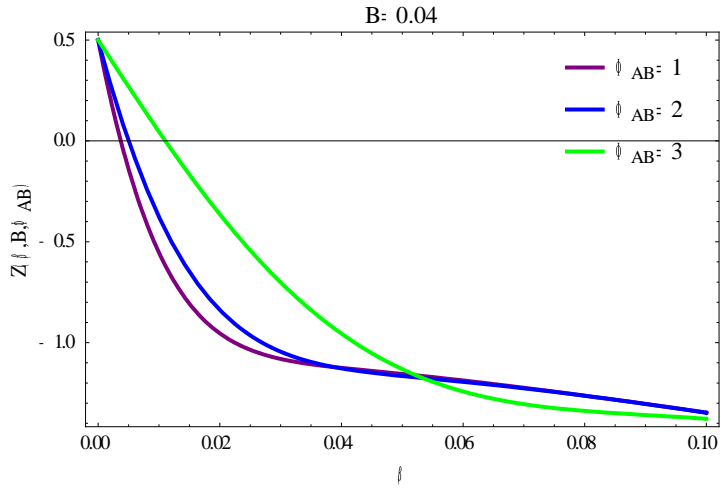

**Figure 8**: Plot of Partition function against $\beta$ for different values AB flux field, $\Phi_{AB}$.

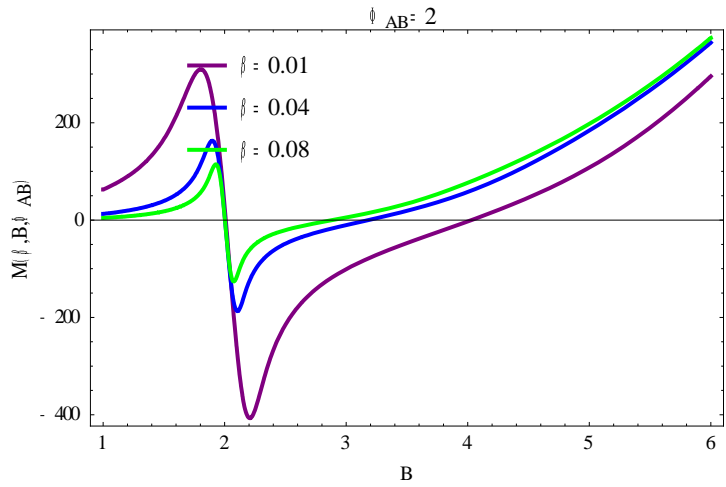

**Figure 9**: Plot of Magnetization against $\vec{B}$ at finite temperature

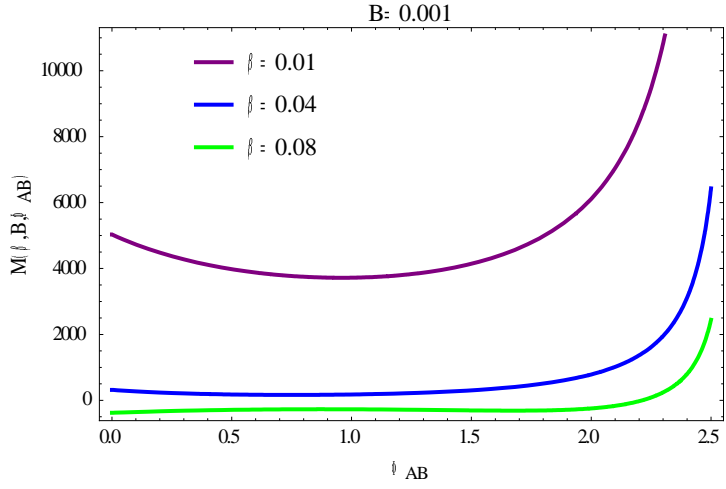

**Figure 10**: Magnetization, $M\left(\vec{B}, \Phi_{AB}, \beta\right)$ against AB flux field, $\Phi_{AB}$ with different $\beta$.

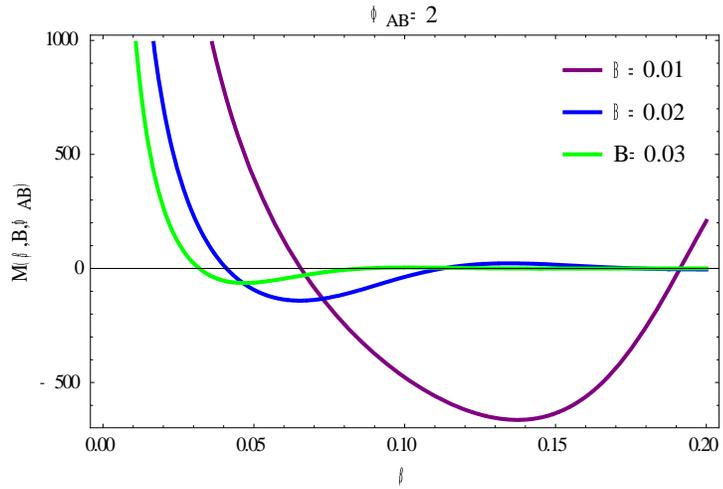

**Figure 11**: Magnetization, $M\left(\vec{B}, \Phi_{AB}, \beta\right)$ against $\beta$ varying $\vec{B}$

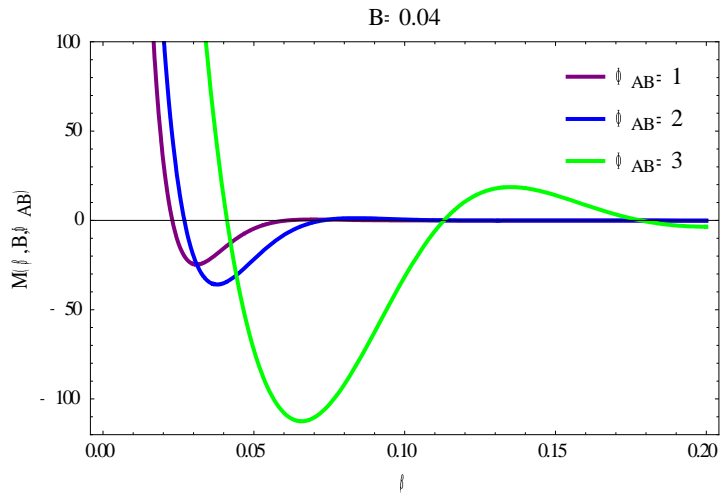

**Figure 12**: Magnetization, $M\left(\vec{B}, \Phi_{AB}, \beta\right)$ against $\beta$ varying $\vec{B}$

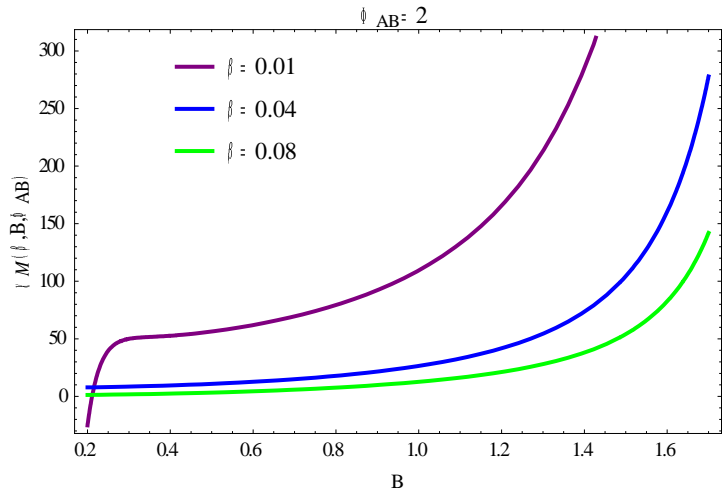

**Figure 13:** Magnetic Susceptibility, $\chi_m\left(\vec{B}, \Phi_{AB}, \beta\right)$ against $\vec{B}$ varying $\beta$

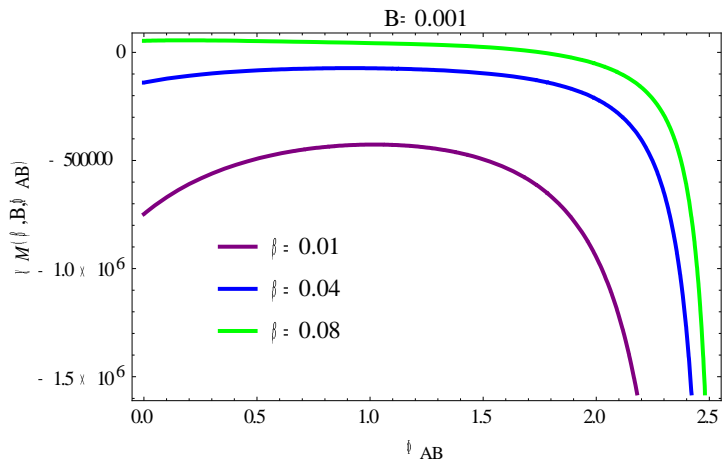

**Figure 14:** Magnetic Susceptibility, $\chi_m\left(\vec{B}, \Phi_{AB}, \beta\right)$ against $\Phi_{AB}$ varying $\beta$

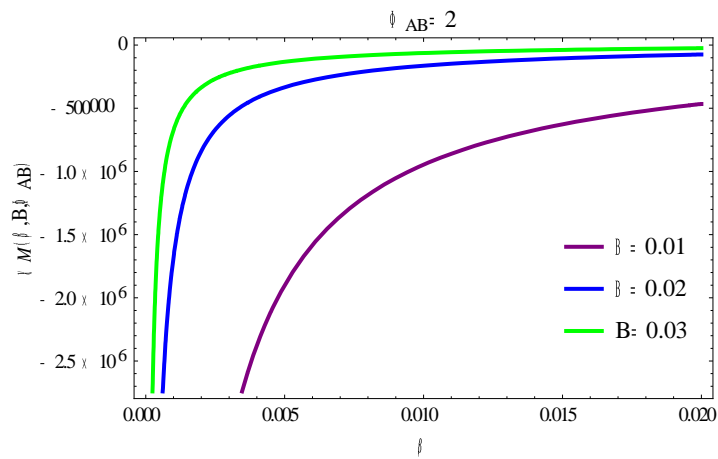

**Figure 15:** Magnetic Susceptibility, $\chi_m\left(\vec{B},\Phi_{AB},\beta\right)$ against $\beta$ varying $\vec{B}$

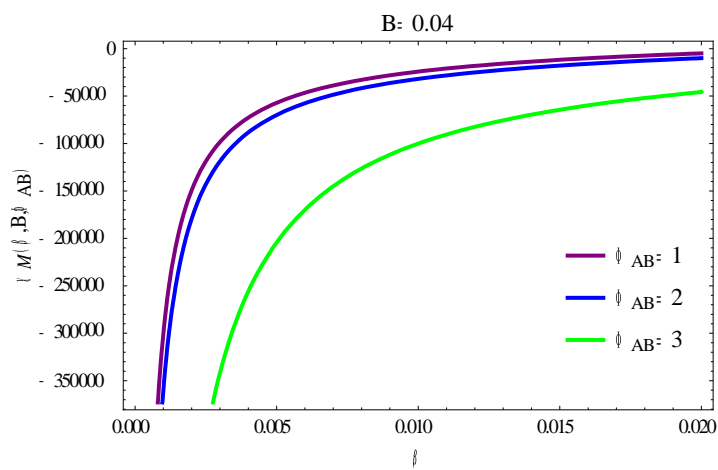

**Figure 16:** Magnetic Susceptibility, $\chi_m\left(\vec{B},\Phi_{AB},\beta\right)$ against $\beta$ varying $\Phi_{AB}$

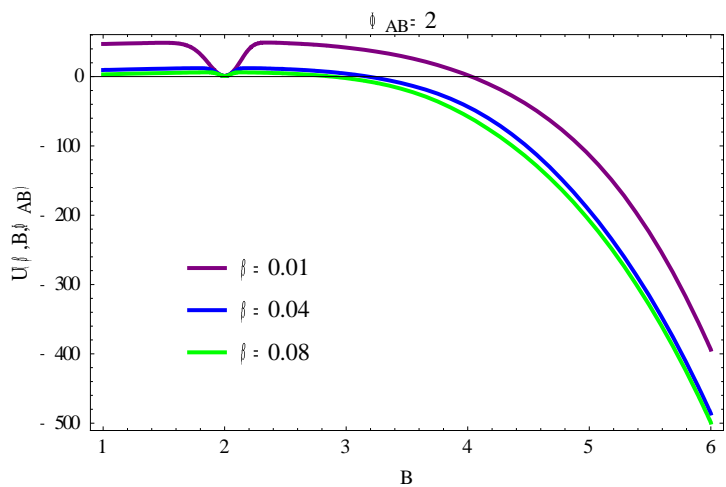

**Figure 17:** Internal Energy, $U(\vec{B}, \Phi_{AB}, \beta)$ against $\vec{B}$ varying $\beta$

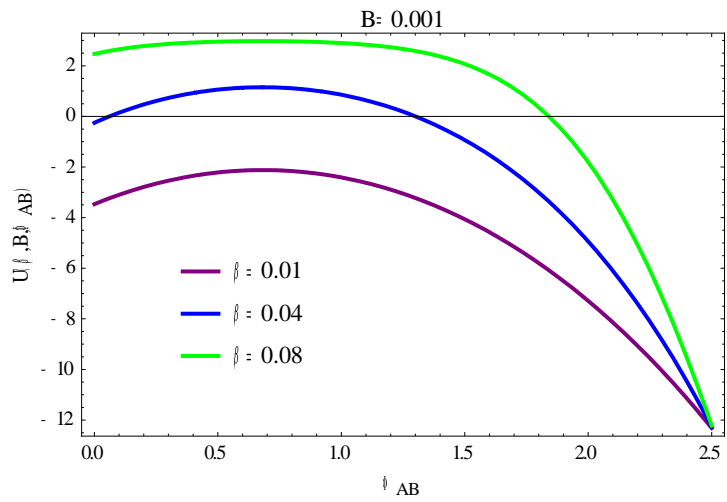

**Figure 18:** Internal Energy, $U(\vec{B}, \Phi_{AB}, \beta)$ against $\Phi_{AB}$ varying $\beta$

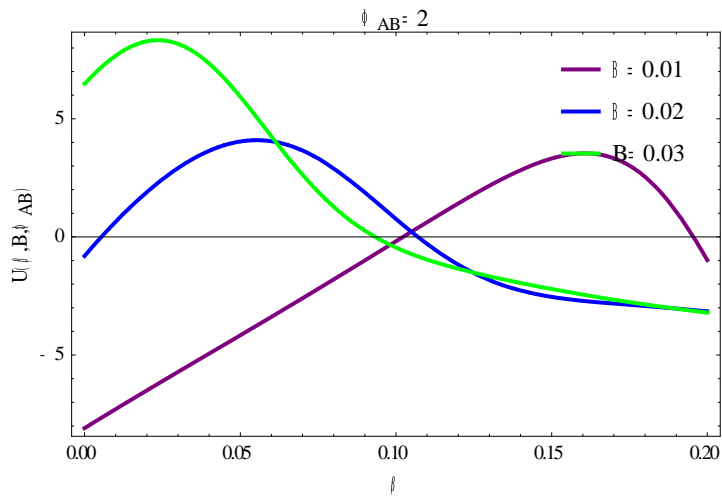

**Figure 19:** Internal Energy, $U(\vec{B}, \Phi_{AB}, \beta)$ against $\beta$ varying $\vec{B}$

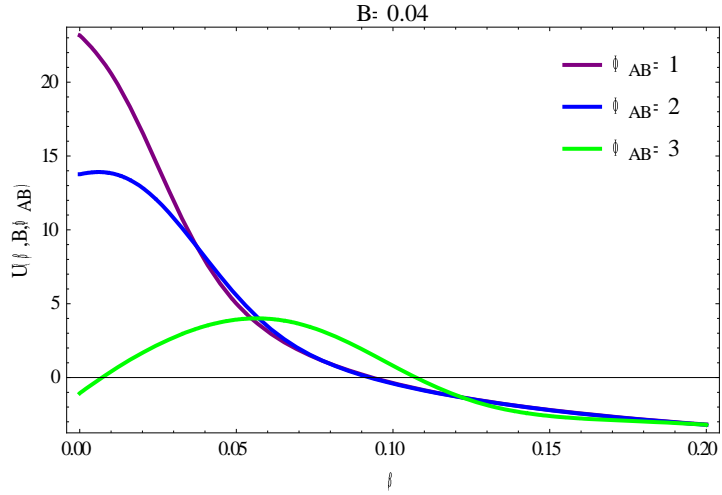

**Figure 20:** Internal Energy, $U\left(\vec{B}, \Phi_{AB}, \beta\right)$ against $\beta$ varying $\Phi_{AB}$

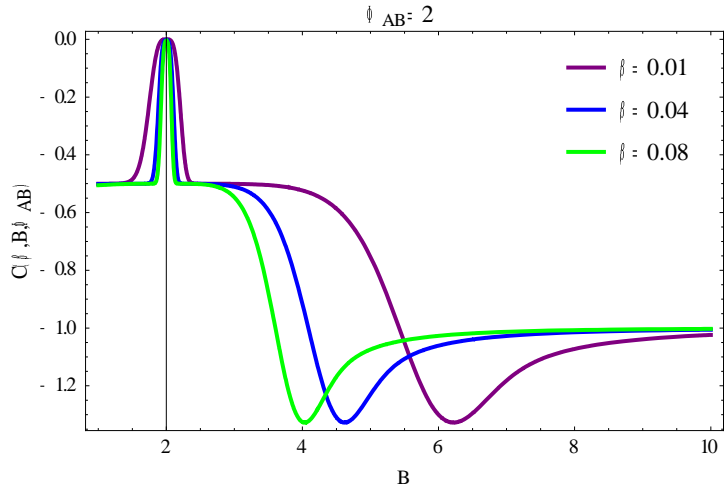

**Figure 21:** Specific Heat Capacity, $C_v\left(\vec{B}, \Phi_{AB}, \beta\right)$ against $\vec{B}$ varying $\beta$

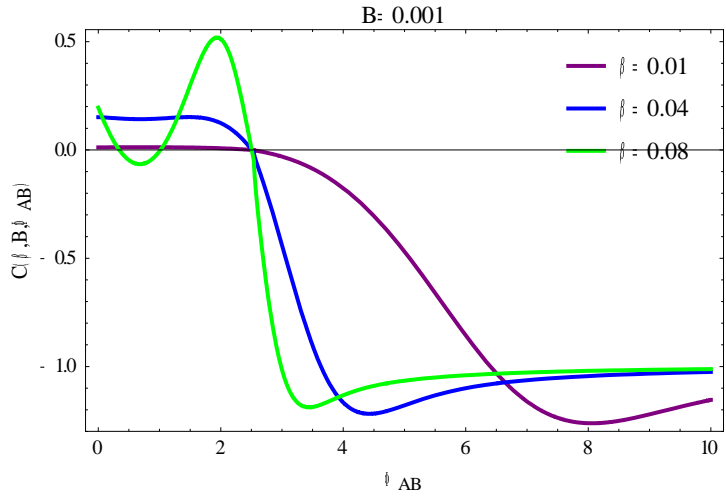

**Figure 22:** Specific heat capacity, $C_v\left(\vec{B},\Phi_{AB},\beta\right)$ against $\Phi_{AB}$ varying $\beta$

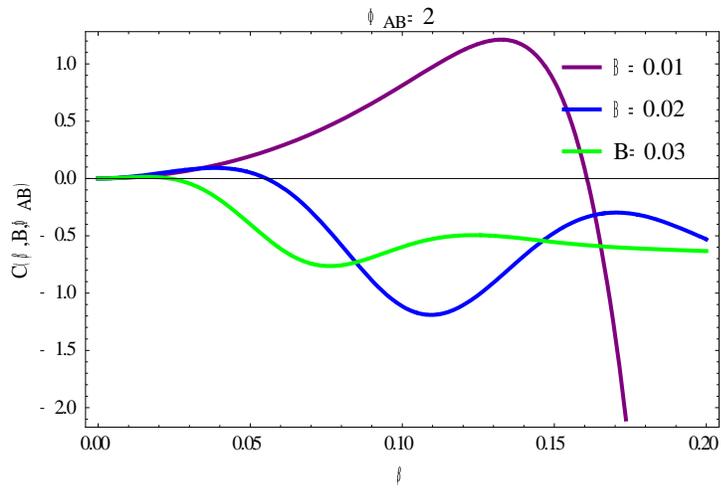

**Figure 23:** Specific heat capacity, $C_v\left(\vec{B},\Phi_{AB},\beta\right)$ against $\beta$ varying $\vec{B}$

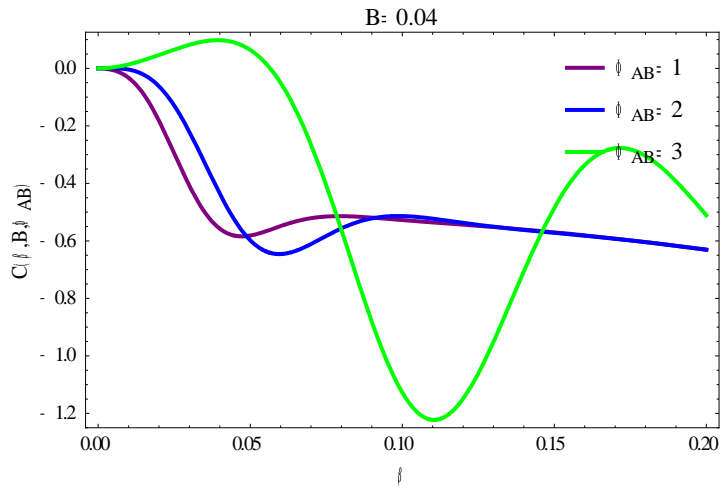

**Figure 24:** Specific heat capacity, $C_v\left(\vec{B}, \Phi_{AB}, \beta\right)$ against $\beta$ varying $\Phi_{AB}$

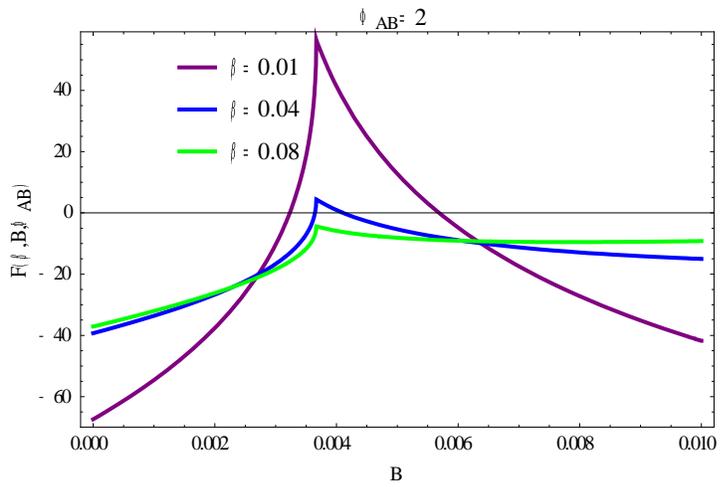

**Figure 25:** Free energy, $F\left(\vec{B}, \Phi_{AB}, \beta\right)$ against $\vec{B}$ varying $\beta$

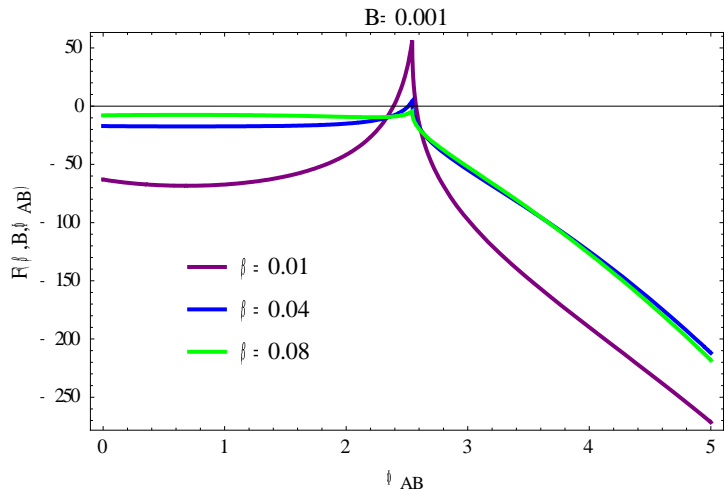

**Figure 26:** Free energy, $F\left(\vec{B}, \Phi_{AB}, \beta\right)$ against $\Phi_{AB}$ varying $\beta$

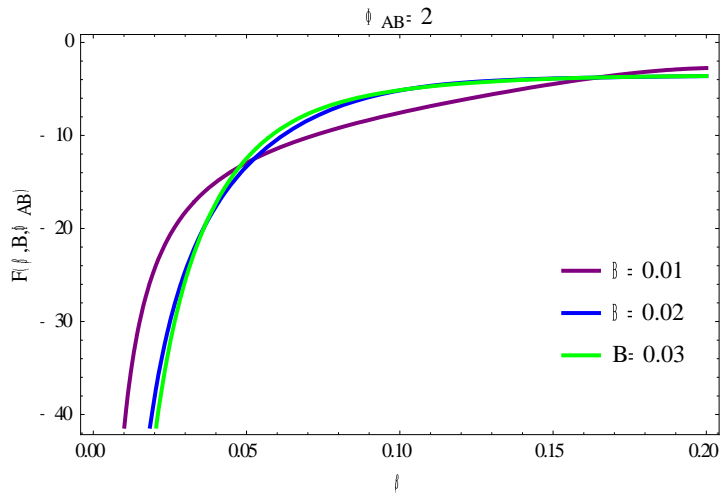

**Figure 27:** Free energy, $F\left(\vec{B}, \Phi_{AB}, \beta\right)$ against $\beta$ varying $\vec{B}$

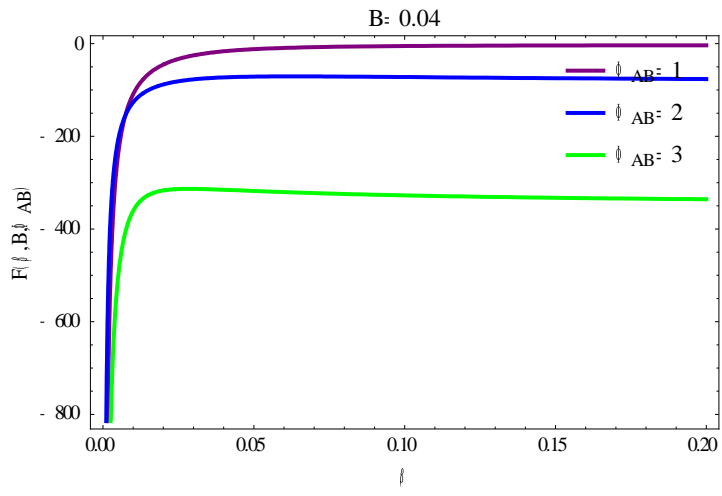

**Figure 28:** Free energy, $F\left(\vec{B}, \Phi_{AB}, \beta\right)$ against $\beta$ varying $\Phi_{AB}$

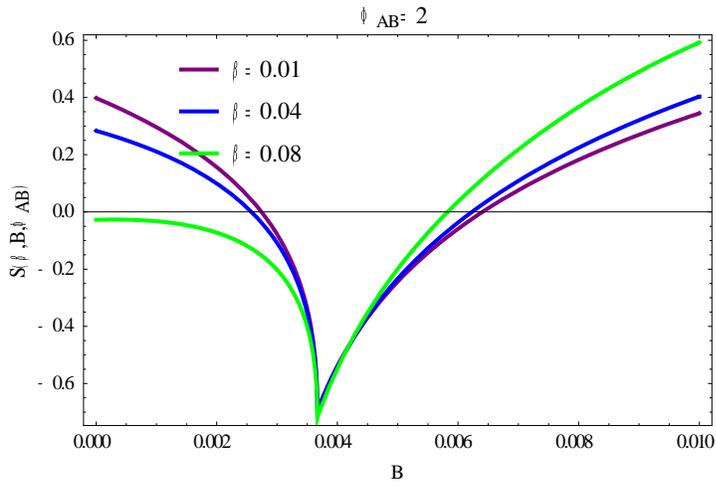

**Figure 29:** Entropy, $S\left(\vec{B}, \Phi_{AB}, \beta\right)$ against $\vec{B}$ varying $\beta$

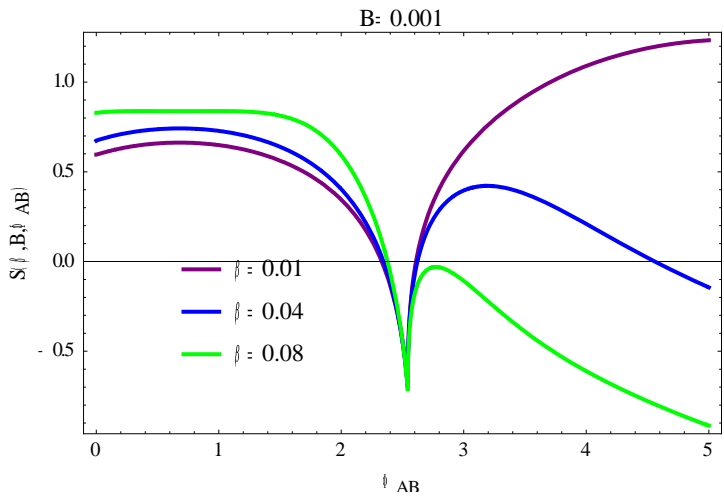

**Figure 30:** Entropy, $S\left(\vec{B}, \Phi_{AB}, \beta\right)$ against $\Phi_{AB}$ varying $\beta$

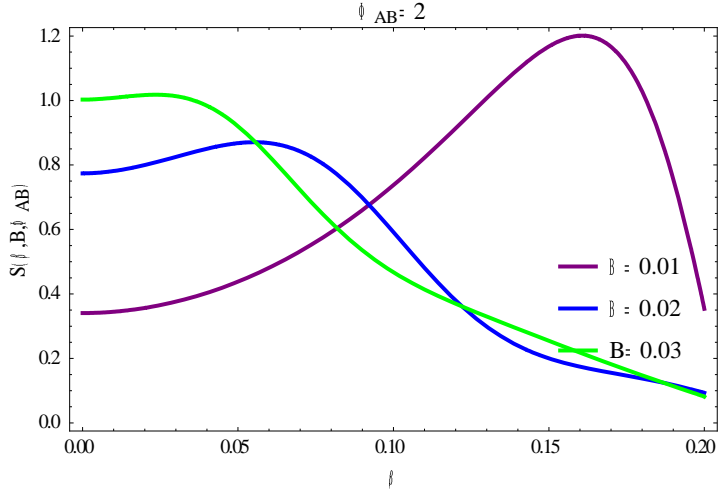

**Figure 31:** Entropy, $S(\vec{B}, \Phi_{AB}, \beta)$ against $\beta$ varying $\vec{B}$

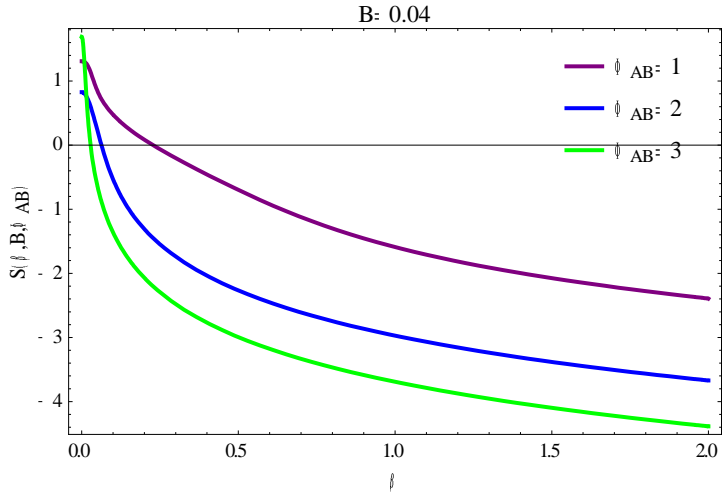

**Figure 32:** Entropy, $S(\vec{B}, \Phi_{AB}, \beta)$ against $\beta$ varying $\Phi_{AB}$

## 6   Conclusion

In this study, the Hellmann potential is examined in the presence of external magnetic and AB-flux fields. The Hamiltonian operator consisting of both fields and the potential is transformed into a second order differential equations. We solve the resulting differential equation via the well-known functional analysis approach to obtain the energy equation and wave function of the system. The effect of the fields on the energy spectra of the system is closely examined. It was found out that the $\vec{B}$ and AB fields removes degeneracy when the screening parameter was $\eta = 0.005$ but when the screening parameter was increased to $\eta = 0.01$, the AB field was found to perform better than the magnetic in its ability to remove degeneracy.

Furthermore, the magnetization and magnetic susceptibility of the system was considered at zero temperature. The system was found to exhibit a paramagnetic behavior $\left(\chi_m(\beta,\Phi_{AB},\vec{B})>0\right)$ and the system also reveals some sort of saturation at large $\vec{B}$. We evaluate the partition function and use it to evaluate other thermodynamic properties of the system such as; magnetic susceptibility, $\chi_m(\vec{B},\Phi_{AB},\beta)$ Helmholtz free energy, $F(\vec{B},\Phi_{AB},\beta)$, entropy, $S(\vec{B},\Phi_{AB},\beta)$, internal energy, $U(\vec{B},\Phi_{AB},\beta)$, and specific heat, $C_v(\vec{B},\Phi_{AB},\beta)$. A comparative analysis of the magnetic susceptibility of the system at zero and finite temperature shows a similarity in the behavior of the system. All thermodynamic properties of Hellmann potential has been thoroughly investigated in presence of both fields. Our research findings could be applied in condensed matter physics, atomic physics and chemical physics.

## REFERENCES

[1] H. Hellmann J. Chem. Phys. **3** (1935) 61

[2] H. Hellmann and W. Kassatotschkin J. Chem. Phys **4** (1936) 324

[3] G. Simons J. Chem. Phys. **55** (1971) 756

[4] H Hellmann, ActaPhysicochim. U.R.S.S. **1**, 913 (1934/1935)

[5] C A Onate, M C Onyeaju, A N Ikot, and O Ebomwonyi, *Eur. Phys. J. Plus* 132 (2017)462

[6] M Hamzavi, K E Thylwe, AARajabi, *Commun. Theor. Phys.***60** (2013) 1
[7] O. J. Oluwadare and K. J. Oyewumi Advance in High Energy Physics **2018** (2018)5214041.
[8] B. H. Yazarloo, H. Mehraban, and H. Hassnabadi, ActaPhysicaPolonica A, **127**(2015)684
[9] O. J. Oluwadare and K. J. Oyewumi, Chin. J. Phys, **55** (2017)2422
[10] J Callaway, P S Laghos, *Phys. Rev.***187** (1969)192.
[11] G McGinn, *J. Chem. Phys.***53**, (1970)3635.
[12] V K Gryaznov, M.V. Zhernokletov, V.N. Zubarev, I.L. Losilevskii, V.E. Tortov, Eh. Eksp. Teor. Fiz. **78** (1980)573
[13] J G Philips, L Kleinmann, *Phys. Rev. A***116** (1959)287
[14] A.J. Hughes, J. Callaway, *Phys. Rev. A***136** (1964)1390
[15] J. Zakrzewski, R. Gebarowski and D. Delande, Phys. Rev. A **54** (1996) 691.

[16] M Eshghi, H Mehraban, and S M Ikhdair Chin. Phys. B **26**, (2017) 060302

[17] M Eshghi, R Sever and S M Ikhdair, Chin. Phys. B **27**, (2018) 020301

[18] M. Eshghi and H. Mehraban. Eur. Phys. J. Plus **132** (2017) 121

[19] R. Khordad, Solid State Sciences **12** (2010) 1253.

[20] S. M. Ikhdair and B. J. Falaye, J. Ass. Arab Univ. Basic & Appl. Sci. **16** (2014) 1.

[21] S. M. Ikhdair, B. J. Falaye and M. Hamzavi Annals of Physics **353** (2015) 282

[22] B. J. Falaye, G. H. Sun, R S. Ortigoz, and S. H. Dong, Phys Rev. E **93** (2016) 053201

[23] M. Aygun, O. Bayrak, I. Boztosun, and Y. Sahin Eur. Phys. J. D **66** (2012) 35

[24] K. J. Oyewumi, E. O. Titiloye, A. B. Alabi and B. J. Falaye Journal of the Nigerian Mathematical Society **35** (2016) 460

[25] N. Ferkous and A. Bounames Commun. Theor. Phys. **59** (2013) 679

[26] A. Çetin. Physica B **523** (2017) 92

[27] F. A. S. Orozco, J. G. A. Ochoa, X. C. Rivas, J. L. C. Figueroa, H. M. M. Carrada. Heliyon **5** (2019) e02224

[28] W. Ebeling and M.I. Sokolov, Statistical Thermodynamics and Stochastic Theory of Nonequilibrium Systems, World Scientific, Singapore (2005) pp. 312.

[29] R.K. Pathria, Statistical Mechanics, Butterworth, Wash- ington (1996).

[30] P.T. Landsberg, Thermodynamics and Statistical Mechanics, Dover, New York (1991).

[31] G. Sukirti, K. Manoj, J.P. Kumar, M. Man, Chin. Phys. B **25**, 056502 (2016)

[32] G.B. Ibragimov, Fizika **34**, 35 (2003)

[33] R. Khordad and H. R. RastegarSedehi J Low Temp Phys **190** (2018) 200

[34] R. Khordad and H.R. RastegarSedehi. Eur. Phys. J. Plus **134** (2019) 133

[35] A. A. Alia, M. K. Elsaid and A. Shaer. Journal of Taibah University for Science **13** (2019) 687

[36] D. A. Baghdasaryan, D. B. Hayrapetyan, E. M. Kazaryan and H. A. Sarkisyan. Physica E: Low-dimensional Systems and Nanostructures **101** (2018) 1

[37] R. Khordad, M.A. Sadeghzadeh, and A. MohamadianJahan-abad Commun. Theor. Phys. **59** (2013) 655

[38] R. L. Greene and C. Aldrich Phys. Rev. A **14** (1976) 2363

[39] S H Dong Factorization Method in Quantum Mechanics (Armsterdam: Springer) (2007)

[40] C. A. Onate, J. O. Ojonubah, A. Adeoti, E. J. Eweh and M. Ugboja. Afr. Rev. Phys. **9** (2014) 0062

[41] O Ebomwonyi, C A Onate, M C Onyeaju and A N Ikot KarbalaInt. J. Mod. **3** (2017) 59

[42] G Kocak, O Bayrak and I Boztosun J. Theor. Comp. Chem. **6** (2007) 893

[43] B. Boyacioglu and A. Chatterjee. Physica E **44** (2012) 1826

[44] U. S. Okorie, E. E. Ibekwe, A. N. Ikot, M. C. OnyeajuandE. O. Chukwuocha, J. Kor. Phys. Soc. **73**(2018)1211.

[45] C-S. Jia, X-T. You, J-Y. Liu, L-H. Zhang, X-L. Peng, Y-T. Wang and L-S. Wei, Chem. Phys. Lett. **717** (2019) 16

[46] C-S. Jia, L-H. Zhang, X-L. Peng, J-X. Luo, Y-L. Zhao, J-Y. Liu, J-J. Guo and L-D. Tang, Chem. Eng. Sci. **202** (2019) 70

[47] Z. Ocak, H. Yanar, M. Saltı and O. Aydoğdu, Chem. Phys. **513**(2018) 252

[48] A. Bera, A. Ghosh and M. Ghosh.J. Magnetism and Magnetic Materials **484** (2019) 391.

[49] U. S. Okorie, A. N. Ikot, M. C. Onyeaju, E. O. Chukwuocha, Rev. Mex. de Fis. **64** (2018) 608.